\documentclass{acm_proc_article-sp-sigmod07}
\usepackage[latin1]{inputenc}
\usepackage{epsfig}
\usepackage{subfigure}
\usepackage{textcomp}
\begin{document}

\title{Perfect Hashing for Data Management Applications}

\numberofauthors{3}

\newcommand{\tabauthors}{Authors}
\newcommand{\tabinstitutions}{Institutions}

\author{
\alignauthor Fabiano C. Botelho\\
       \affaddr{Dept. of Computer Science}\\
       \affaddr{Federal Univ. of Minas Gerais}\\
       \affaddr{Belo Horizonte, Brazil}\\
       \email{fbotelho@dcc.ufmg.br}
\
\alignauthor Rasmus Pagh \\
       \affaddr{Computational Logic and Algorithms Group}\\
       \affaddr{IT Univ. of Copenhagen}\\
       \affaddr{Copenhagen, Denmark}\\
       \email{pagh@itu.dk}
\
\alignauthor Nivio Ziviani\\
       \affaddr{Dept. of Computer Science}\\
       \affaddr{Federal Univ. of Minas Gerais}\\
       \affaddr{Belo Horizonte, Brazil}\\
       \email{nivio@dcc.ufmg.br}
}



\maketitle

\begin{abstract}
Perfect hash functions can potentially be used to compress data in
connection with a variety of data management tasks.
Though there has been considerable work on how to construct
good perfect hash functions, 
there is a gap between theory and practice 
among all previous methods on minimal perfect hashing.
On one side, there are good theoretical results without
experimentally proven practicality for large key sets. 
On the other side, there are 
the theoretically analyzed time and space usage
algorithms that assume that truly random hash functions are available 
for free, which is an unrealistic assumption.
In this paper we attempt to bridge this gap between theory and
practice, using a number of techniques from the literature to obtain
a novel scheme that is theoretically well-understood and at the same time 
achieves an order-of-magnitude increase in performance 
compared to previous ``practical'' methods. This improvement
comes from a combination of a novel, theoretically optimal perfect hashing
scheme that greatly simplifies previous methods, 
and the fact that our algorithm is designed to make good use
of the memory hierarchy.
We demonstrate the scalability of our algorithm by considering a set of over one billion 
URLs from the World Wide Web of average length 64, for which we construct a minimal 
perfect hash function on a commodity PC in a little more than 1 hour. 
Our scheme produces minimal perfect hash functions using slightly more than 3 bits per key. 
For perfect hash functions in the range $\{0,\dots,2n-1\}$ the space usage drops to 
just over 2 bits per key (i.e., one bit 
more than optimal for representing the key). 
This is significantly below of what has been achieved previously for 
very large values of $n$. 
\end{abstract}

\def\cG{{\mathcal G}}
\def\crit{{\rm crit}}
\def\ncrit{{\rm ncrit}}
\def\scrit{{\rm scrit}}
\def\bedges{{\rm bedges}}
\def\ZZ{{\mathbb Z}}
\def\BSmax{\mathit{BS}_{\mathit{max}}}
\def\Bi{\mathop{\rm Bi}\nolimits}

\section{Introduction}
\label{sec:intro}

Some types of databases are updated only rarely, typically by periodic batch updates. 
This is true, for example, for most data warehousing applications (see~\cite{s05} 
for more examples and discussion). 
In such scenarios it is possible to improve query performance by creating 
very compact representations of keys by minimal perfect hash functions.
In applications where the set of keys is fixed for a long period of time
the construction of a minimal perfect hash function can be done as part of
the preprocessing phase.
For example, On-Line Analytical Processing (OLAP) applications use extensive preprocessing 
of data to allow very fast evaluation of certain types of queries.

{\em Perfect hashing} is a space-efficient way of creating compact representation
for a static set $S$ of $n$ keys.
For applications with successful searches, the representation of a key
$x \in S$ is simply the value $h(x)$, where $h$ is a perfect hash function for
the set $S$ of values considered.
The word ``perfect'' refers to the fact that the function will map the elements 
of $S$ to unique values (is identity preserving). 
{\em Minimal perfect hash function} (MPHF) produces values that are integers in
the range $[0, n-1]$, which is the smallest possible range.
It is known that $O(n)$ bits suffice to store a minimal perfect hash function, 
and there are theoretical results that use around $1.4427 n$ bits, 
asymptotically for large $n$~\cite{ht01}. 

We now present some examples where minimal perfect hash functions 
have successfully been applied to:
\begin{itemize}
\item
A perfect hash function can be used to implement a data structure
with the same functionality as a Bloom filter~\cite{ppr05}.
In many applications where a set $S$ of elements is to be stored,
it is acceptable to include in the set some false 
positives\footnote{False positives are elements that appear 
to be in $S$ but are not.}
with a small probability by storing a signature for each perfect
hash value.
This data structure requires around 30\% less space usage when compared
with Bloom filters, plus the space for the perfect hash function.
Bloom filters have applications in distributed databases and data mining
(association rule mining~\cite{cl05, clc06}).
\item
Perfect hash functions have also been used to speed up the partitioned 
hash-join algorithm presented in \cite{mbk00}. 
By using a perfect hash function to reduce the
targeted hash-bucket size from 4 tuples to just 1 tuple they have avoided
following the bucket-chain during hash-lookups that  causes too many cache
and translation lookaside buffer (TLB) misses.
\item
Suppose there is a composite foreign key to a table T of size n.
Then the size of the key needed in T will typically be larger than
log n. For example, suppose tuples of R contain geographical
coordinates that are used as foreign key references. Replacing
the coordinates with a surrogate key may be a bad choice if common
queries on R involve conditions on the coordinates, as a join would
be required to retrieve the coordinates. In general, if we have a natural
foreign key that carries information relevant for queries, we can avoid
the cost of additionally storing a surrogate key.
\end{itemize}

The perfect hash function used depends on the set $S$ of distinct attribute values that occur.
It is known that maintaining a perfect hash function dynamically under insertions into $S$ is 
only possible using space that is super-linear in $n$~\cite{dictionariis}. 
However, in this paper we consider the case where $S$ is fixed, and construction of a 
perfect hash function can be done as part of the preprocessing of data (e.g., in a data warehouse).
To the best of our knowledge, previously suggested perfect hashing methods have not been 
even close to a space usage of $1.4427 n$ bits for realistic data sizes. 
Second, all previous methods suffer from either an incomplete theoretical understanding 
(so there is no guarantee that it works well on a given data set) or seems impractical 
due to a very intricate and time-consuming evaluation procedure. 

In this paper we present a scalable algorithm that produces minimal perfect hash 
functions using slightly more than 3 bits per key. Also, if we are happy with values in the 
range $\{0,\dots,2n-1\}$ (i.e., using one bit more than optimal for representing 
the surrogate key) the space usage drops to just over 2 bits per key. 
This is significantly below of what has been achieved previously for imaginable 
values of $n$. We demonstrate the scalability of our algorithm by considering a 
set of over one billion strings (URLs from the world wide web of average length 64), 
for which we construct a minimal perfect hash function on a commodity PC in a little 
more than 1 hour. 
This is an order of magnitude faster than previous methods. 
If we use the range $\{0,\dots,2n-1\}$, the space for the perfect hash function is 
less than 324 MB, and we still get hash values that can be represented in a 32 bit word. 
Thus we believe our MPHF method might be quite useful for a number of
current and practical data management problems.

\section{Notation}

Suppose~$U$ is a universe of \textit{keys} of size $u$.
Let $h:U\to M$ be a {\em hash function} that maps the keys from~$U$
to a given interval of integers $M=[0,m-1]=\{0,1,\dots,m-1\}$.
Let~$S\subseteq U$ be a set of~$n$ keys from~$U$, where $ n \ll u$.
Given a key~$x\in S$, the hash function~$h$ computes an integer in
$[0,m-1]$.
A perfect hash function (PHF) maps a set $S$ of $n$ keys from $U$ into a set of $m$ integer 
numbers without collisions, where $m$ is greater than or equal to $n$. 
If $m$ is equal to $n$, the function is called minimal (MPHF). 

\section{Related work}
\label{sec:relatedprevious-work}


There is a gap between theory and practice 
among minimal perfect hashing methods.
On one side, there are good theoretical results without
experimentally proven practicality for large key sets. We will argue below
that these methods are indeed not practical.
On the other side, there are two categories of practical algorithms:
the theoretically analyzed time and space usage
algorithms that assume truly random hash functions for their methods,
which is an unrealistic assumption,
and the algorithms that present only empirical evidences.
The aim of this section is to discuss the existent gap
among these three types of algorithms available in the literature.

\subsection{Theoretical results}


In this section we review some of the most important theoretical results on minimal 
perfect hashing. 
For a complete survey until 1997 refer to Czech, Havas and Majewski~\cite{chm97}.

Fredman, Koml\'os and Szemer\'edi~\cite{fks84s} proved, using 
a counting argument, that at least 
$n\log e + \log \log u - O(\log n)$ bits 
are required to represent a MPHF,
provided that $u \geq n^{2+\jmath}$ for some $\jmath > 0$ (an easier proof was given 
by Radhakrishnan~\cite{r92}).
Mehlhorn~\cite{m84} has made this bound almost tight
by providing an algorithm that constructs a MPHF that
can be represented with at most $n\log e + \log \log u + O(\log n)$ bits.
However, his algorithm is far away from practice because its
construction and evaluation time is exponential on $n$ 
(i.e., $n^{\theta(n e^n u \log u)}$).

Schmidt and A. Siegel \cite{ss90} have proposed the 
first algorithm for constructing a MPHF with constant 
evaluation time and description size $O(n + \log \log u)$ bits.
Nevertheless, 
the scheme is hard to implement and the constants 
associated with the MPHF 
storage are prohibitive. For a set of $n$ keys, at least $29n$ bits are used,
which means that the space usage is similar in practice to schemes using $n\log n$ bits.

More recently, Hagerup and Tholey~\cite{ht01} have come up 
with the best theoretical result we know of. 
The MPHF obtained can be evaluated in $O(1)$ time and 
stored in $n\log e + \log \log u + O(n(\log \log n)^2/\log n + \log \log \log u)$
bits. 
The construction time is $O(n + \log \log u)$ using $O(n)$
computer words of the Fredman, Koml\'os and Szemer\'edi~\cite{FKS84} model of computation.
In this model, also called the {\em Word RAM\/} model, an element of the universe~$U$ fits
into one machine word, and arithmetic operations and memory accesses have unit
cost.  
In spite of its theoretical importance, the Hagerup and Tholey~\cite{ht01} algorithm
is not practical as well, as it emphasizes asymptotic space complexity only. (It is also very
complicated to implement, but we will not go into that.)
For $n<2^{150}$ the scheme is not even defined, as it relies on splitting the key set
into buckets of size $\hat{n}\leq \log n/(21\log\log n)$. Even if we fix this by letting the
bucket size be at least 1, then for $n<2^{300}$ buckets of size one will be used, which
means that the space usage will be at least $(3\log\log n+\log 7)\, n$ bits. For a set of
a billion keys, this is more than 17 bits per element. Since $2^{300}$ exceeds the number
of atoms in the known universe, it is safe to conclude that the Hagerup-Tholey algorithm
will not be space efficient in practical situations.



\subsection{Practical results assuming full randomness}


Let us now describe the main practical results
analyzed with the unrealistic
assumption that truly random hash functions are available for free.

Fox et al.~\cite{fhcd92} created the first scheme 
with good average-case performance for large datasets, i.e., $n\approx 10^6$.
They have designed two algorithms, the first one generates a
MPHF that can be evaluated in $O(1)$ time and stored in $O(n \log n)$ bits.
The second algorithm uses quadratic hashing and adds branching based on a table of 
binary values to get a MPHF that can be evaluated in $O(1)$ time and 
stored in $c(n + 1/\log n)$ bits. 
They argued that $c$ would be typically lower than 5, however,
it is clear from their experimentation that $c$ grows with $n$ and 
they did not discuss this.
They claimed that their algorithms would 
run in linear time, but,
it is shown in~\cite[Section 6.7]{chm97} that the algorithms have exponential 
running times in the worst case, although the worst case has small probability of 
occurring. Fox, Chen and Heath~\cite{fch92} improved the above result
to get a function that can be stored in $cn$ bits. 
The method uses four truly random hash functions 
$h_{10}:S \to [0, n-1]$,
$h_{11}:[0, p_1 - 1] \to [0, p_2-1]$,
$h_{12}:[p_1, n - 1] \to [p_2, b-1]$ and
$h_{20}:S \times \{0,1\}\to [0, n-1]$
to construct a MPHF that has the following form:
{\small
\begin{eqnarray*}
h(x) &=& (h_{20}(x, d) + g(i(x)))\bmod n \\
i(x) &=&
\begin{cases}
h_{11}\circ h_{10}(x) & \text{if $h_{10}(x) < p_1$}\\
h_{12}\circ h_{10}(x) & \mbox{otherwise.}
\end{cases}
\end{eqnarray*}
}where $p_1=0.6n$ and $p_2 = 0.3n$ were experimentally determined,
and $\lceil b = cn/(\log n + 1) \rceil$.
Again $c$ is only established for small values of $n$.
It could very well be that $c$ grows with $n$.
So, the limitation of the three algorithms is that no guarantee on the size 
of the resulting MPHF is provided.

The family of algorithms proposed by Czech et al~\cite{mwhc96} 
uses the same assumption to construct order preserving
MPHF. A perfect hash function $h$ is
\textit{order preserving} if the keys in~$S$ are arranged in some given order
and~$h$ preserves this order in the hash table.
The method uses two truly random hash functions
$h_1(x): S \to cn$ and  $h_2(x): S \to cn$ to
generate MPHFs in the following form: $h(x) = (g[h_1(x)] + g[h_2(x)]\bmod n$
where $c > 2$.
The resulting MPHFs can be evaluated in $O(1)$ time and stored in
$O(n \log n)$ bits (that is optimal for an order preserving MPHF).
The resulting MPHF is generated in expected $O(n)$ time.
Botelho, Kohayakawa and Ziviani~\cite{bkz05} improved
the space requirement at the expense of generating functions in the same form
that are not order preserving. 
Their algorithm is also linear on $n$, but runs faster 
than the ones by Czech et al~\cite{mwhc96} and 
the resulting MPHF are stored using half of the space because $c\in[0.93,1.15]$. 
However, the  resulting MPHFs still need $O(n \log n)$ bits to be stored. 

Since the space requirements for truly random hash functions makes them unsuitable for
implementation, one has to settle for a more realistic setup. 
The first step in this direction was given by Pagh~\cite{p99}. 
He proposed a family of randomized algorithms for
constructing MPHFs
of the form $h(x) = (f(x) + d[g(x)]) \bmod n$,
where $f$ and $g$ are chosen from a family of universal hash functions and $d$ is a set of
displacement values to resolve collisions that are caused by the function $f$.
Pagh identified a set of conditions concerning $f$ and $g$ and showed
that if these conditions are satisfied, then a minimal perfect hash
function can be computed in expected time $O(n)$ and stored in
$(2+\epsilon)n\log n$ bits.

Dietzfelbinger and Hagerup~\cite{dh01} improved the algorithm proposed in~\cite{p99},
reducing from $(2+\epsilon)n\log n$ to $(1+\epsilon)n\log n$ the number of bits
required to store the function, but in their approach~$f$ and~$g$ must
be chosen from a class
of hash functions that meet additional requirements.


\subsection{Empirical results}

In this section we discuss results that present only empirical evidences
for specific applications.
Lefebvre and Hoppe~\cite{lh06} have recently designed MPHFs that require 
up to 7 bits per key to be stored and are tailored to represent 
sparse spatial data.
In the same trend,
Chang, Lin and Chou~\cite{cl05, clc06} have designed MPHFs
tailored for mining association rules 
and traversal patterns in data mining techniques.

\subsection{Differences between our results and previous results}


In this work we propose an algorithm that is 
theoretically well-understood and achieves an
order-of-magnitude increase in the performance on a commodity PC
compared to previous ``practical'' methods. 
To the best of our knowledge our algorithm 
is the first one that demonstrates the capability of 
generating MPHFs for sets in the order
of billions of keys, and the generated functions  
require less than 4 bits per key to be stored.
This increases one order of magnitude in the size of the greatest 
key set for which a MPHF was obtained in the literature~\cite{bkz05}.
This improvement
comes mainly from the fact that our method is designed to make good use
of the memory hierarchy. We need $O(N)$ computer words,
where $N \ll n$, for the construction process.
Notice that both space usage 
for representing the MPHF and the construction time are carefully proven.
Additionally, our scheme is much simpler than previous theoretical well-founded schemes.

\section{The algorithm}
\label{sec:new-algorithm}


Our algorithm uses the well-known idea of partitioning the key set into a number of small sets\footnote{Used in e.g.~the perfect hash function constructions of Schmidt and Siegel~\cite{ss90} and Hagerup and Tholey~\cite{ht01}, for suitable definition of ``small''.} (called ``buckets'') using a hash function $h_0$. Let $B_i = \{ x\in S \;|\; h_0(x)=i\}$ denote the $i$th bucket. If we define {\em offset}$[i]=\sum_{j=0}^{i-1} |B_i|$ and let $p_i$ denote a MPHF for $B_i$ then clearly 
\begin{equation}
p(x)= p_i(x) + \text{\em offset}[h_0(x)]	
\end{equation}
is a MPHF for the whole set $S$. 
Thus, the problem is reduced to computing and storing the offset array, as well as the MPHF for each bucket.

\begin{figure}[ht]
\centering
\begin{picture}(0,0)%
\includegraphics{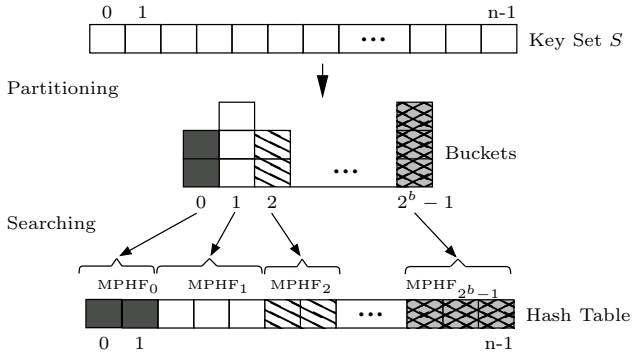}%
\end{picture}%
\setlength{\unitlength}{4144sp}%
\begingroup\makeatletter\ifx\SetFigFont\undefined%
\gdef\SetFigFont#1#2#3#4#5{%
  \reset@font\fontsize{#1}{#2pt}%
  \fontfamily{#3}\fontseries{#4}\fontshape{#5}%
  \selectfont}%
\fi\endgroup%
\begin{picture}(3704,2091)(1426,-5161)
\put(4563,-3329){\makebox(0,0)[lb]{\smash{{\SetFigFont{7}{8.4}{\familydefault}{\mddefault}{\updefault}Key Set $S$}}}}
\put(2009,-3160){\makebox(0,0)[lb]{\smash{{\SetFigFont{7}{8.4}{\familydefault}{\mddefault}{\updefault}0}}}}
\put(2221,-3160){\makebox(0,0)[lb]{\smash{{\SetFigFont{7}{8.4}{\familydefault}{\mddefault}{\updefault}1}}}}
\put(4315,-3160){\makebox(0,0)[lb]{\smash{{\SetFigFont{7}{8.4}{\familydefault}{\mddefault}{\updefault}n-1}}}}
\put(1992,-5146){\makebox(0,0)[lb]{\smash{{\SetFigFont{7}{8.4}{\familydefault}{\mddefault}{\updefault}0}}}}
\put(2204,-5146){\makebox(0,0)[lb]{\smash{{\SetFigFont{7}{8.4}{\familydefault}{\mddefault}{\updefault}1}}}}
\put(4298,-5146){\makebox(0,0)[lb]{\smash{{\SetFigFont{7}{8.4}{\familydefault}{\mddefault}{\updefault}n-1}}}}
\put(4546,-4977){\makebox(0,0)[lb]{\smash{{\SetFigFont{7}{8.4}{\familydefault}{\mddefault}{\updefault}Hash Table}}}}
\put(1981,-4786){\makebox(0,0)[lb]{\smash{{\SetFigFont{5}{6.0}{\familydefault}{\mddefault}{\updefault}MPHF$_0$}}}}
\put(2521,-4786){\makebox(0,0)[lb]{\smash{{\SetFigFont{5}{6.0}{\familydefault}{\mddefault}{\updefault}MPHF$_1$}}}}
\put(3016,-4786){\makebox(0,0)[lb]{\smash{{\SetFigFont{5}{6.0}{\familydefault}{\mddefault}{\updefault}MPHF$_2$}}}}
\put(3826,-4786){\makebox(0,0)[lb]{\smash{{\SetFigFont{5}{6.0}{\familydefault}{\mddefault}{\updefault}MPHF$_{2^b - 1}$}}}}
\put(1441,-3616){\makebox(0,0)[lb]{\smash{{\SetFigFont{7}{8.4}{\familydefault}{\mddefault}{\updefault}Partitioning}}}}
\put(1441,-4426){\makebox(0,0)[lb]{\smash{{\SetFigFont{7}{8.4}{\familydefault}{\mddefault}{\updefault}Searching}}}}
\put(2570,-4301){\makebox(0,0)[lb]{\smash{{\SetFigFont{7}{8.4}{\familydefault}{\mddefault}{\updefault}0}}}}
\put(2782,-4301){\makebox(0,0)[lb]{\smash{{\SetFigFont{7}{8.4}{\familydefault}{\mddefault}{\updefault}1}}}}
\put(2996,-4301){\makebox(0,0)[lb]{\smash{{\SetFigFont{7}{8.4}{\familydefault}{\mddefault}{\updefault}2}}}}
\put(4060,-4006){\makebox(0,0)[lb]{\smash{{\SetFigFont{7}{8.4}{\familydefault}{\mddefault}{\updefault}Buckets}}}}
\put(3776,-4301){\makebox(0,0)[lb]{\smash{{\SetFigFont{7}{8.4}{\familydefault}{\mddefault}{\updefault}$2^b - 1$}}}}
\end{picture}%
\caption{Main steps of our algorithm}
\label{fig:new-algo-main-steps}
\end{figure}

Figure~\ref{fig:new-algo-main-steps} illustrates the two steps of the
algorithm: {\it the partitioning} step and {\it the searching step}.
The partitioning step takes a key set $S$ and uses a hash function 
$h_0$ to partition~$S$ into $2^b$ buckets.
The searching step generates a MPHF $p_i$ for each bucket $i$, 
$0 \leq i \leq 2^b - 1$ and computes the offset array.
From now on the algorithm used to compute the MPHF of each bucket is referred 
to as {\it internal algorithm} and the whole scheme is referred to as 
{\it external algorithm}.

We will choose $h_0$ such that it has values in $\{0,1\}^b$, for some integer $b$. Since the offset array holds $2^b$ entries of at least $\log n$ bits we want $2^b$ to be less than around $n / \log n$,
making the space used for the offset array negligible. On the other hand, to allow efficient implementation of the functions $p_i$ we impose an upper bound $\ell$ on the size of any bucket.
We will describe later how to choose $h_0$ such that this upper bound holds.

To create the MPHFs $p_i$ we could choose from a number of alternatives, emphasizing either space usage, construction time, or evaluation time. We show that all methods based on the assumption of truly random hash functions can be made to work, with explicit and provably good hash functions. For the experiments we have implemented the algorithm described in Section~\ref{sec:simple-mphf}.
Since this computation is done on a small set, we can expect nearly all memory accesses to be ``cache hits''. We believe that this is the main reason why our method performs better than previous ones that access memory in a more ``random'' fashion.


We consider the situation in which the set of all keys may not fit in the internal memory and has to be written on disk. Our external algorithm first scans the list of keys and computes the hash function values that will be needed later on in the algorithm. These values will (with high probability) distinguish all keys, so we can discard the original keys. It is well known that hash values of at least $2\log n$ bits are required to make this work. Thus, for sets of a billion keys or more we cannot expect the list of hash values to fit in the internal memory of a standard PC.

To form the buckets we sort the hash values of the keys according to the value of $h_0$. Since we are interested in scalability to large key sets, this is done using an implementation of an external memory mergesort~\cite{lg98}. 
If the merge sort works in two phases, which is the case for all reasonable parameters, the total work on the disk consists of reading the keys, plus writing and reading the hash function values once. Since the $h_0$ hash values are relatively small (less than 15 decimal digits) 
we can use radix sort to do the internal memory sorting.

The detailed description of the external algorithm is presented in 
Section~\ref{sec:impl.-of-the-algorithm}.
The presentation of the internal algorithm used to compute the MPHF of 
each bucket is presented in
Section~\ref{sec:simple-mphf}.
The internal algorithm uses two hash functions $h_{i1}$ and $h_{i2}$
to compute a MPHF $p_i$. These hash functions as well as 
the hash function $h_0$ used in the partitioning step of the external
algorithm are described in Section~\ref{sec:hashfunctions}.

\subsection{The external algorithm}
\label{sec:impl.-of-the-algorithm}
In this section we are going to present
the implementation of the two-step external memory based algorithm
and the values of the parameters related to the algorithm.
The algorithm is essentially a 
two-phase multi-way merge sort with some nuances 
to make it work in linear time.

The partitioning step performs two important tasks.
First, the variable-length keys are mapped to
128-bit strings by using the linear hash function $h'$ presented in 
Section~\ref{sec:hashfunctions}. That is, the variable-length
key set $S$ is mapped to a fixed-length key set $F$. 
Second, the set $S$ of $n$ keys is partitioned into $2^b$ buckets, 
where  $b$ is a suitable parameter chosen to guarantee
that each bucket has at most $\ell = 256$ keys with high probability 
(see Section~\ref{sec:hashfunctions}).
We have two reasons for choosing $\ell = 256$. 
The first one is to keep the buckets size small enough 
to be represented by 8-bit integers. 
The second one is to allow the memory accesses during the MPHF evaluation
to be done in the cache most of the time.
Figure~\ref{fig:partitioningstep} presents the partitioning step algorithm.
\begin{figure}[h]
\hrule 
\hrule 
\vspace{2mm}
\begin{tabbing}
aa\=type booleanx \==  (false, true); \kill
\> $\blacktriangleright$ Let $\beta$ be the size in bytes of the fixed-length key \\ 
\> ~~~ set $F$ \\ 
\> $\blacktriangleright$ Let $\mu$ be the size in bytes of an a priori reserved \\
\> ~~~ internal memory area \\ 
\> $\blacktriangleright$ Let $N = \lceil \beta/\mu \rceil$ be the number of key blocks that \\
\> ~~~ will be read from disk into an internal memory area \\
\> $1.$ {\bf for} $j = 1$ {\bf to} $N$ {\bf do} \\
\> ~~ $1.1$ Read a key block $S_j$ from disk (one at a time) \\
\> ~~~~~~~ and store $h'(x)$, for each $x\in S_j$, into $\cal{B}$$_j$, \\
\> ~~~~~~~ where $|\cal{B}$$_j| = \mu$ \\
\> ~~ $1.2$ Cluster $\cal{B}$$_j$ into $2^b$ buckets using an indirect radix\\
\> ~~~~~~~ sort algorithm that takes $h_0(x)$ for $x\in S_j$ as \\
\> ~~~~~~~ sorting key(i.e, the $b$ most significant bits of $h'(x)$)\\
\> ~~ $1.3$ Dump $\cal{B}$$_j$ to the disk into File $j$
\end{tabbing}
\hrule 
\hrule 
\vspace{-1.0mm}
\caption{Partitioning step}
\label{fig:partitioningstep}
\vspace{-3.0mm}
\end{figure}

The critical point in Figure~\ref{fig:partitioningstep} 
that allows the partitioning step to work in linear time 
is the internal sorting algorithm.
We have two reasons to choose radix sort.
First, it sorts each key block $\cal{B}$$_j$ in linear time,
since keys are short integer numbers (less than 15 decimal digits).
Second, it just needs $O(|\cal{B}$$_j|)$ words of extra memory so that 
we can control the memory usage independently of the number of keys in $S$.


At this point one could ask: why not to use the well known 
replacement selection algorithm to build files larger than
the internal memory area size? 
The reason is that the radix sort algorithm 
sorts a block $\cal{B}$$_j$ in time $O(|\cal{B}$$_j|)$
while the replacement selection algorithm 
requires $O(|\cal{B}$$_j| \log |\cal{B}$$_j|)$. 
We have tried out both versions and 
the one using the radix sort algorithm outperforms
the other. A worthwhile optimization we have used 
is the last run optimization proposed by Larson and Graefe~\cite{lg98}.
That is, the last block is kept in memory instead of dumping it to disk
to be read again in the second step of the algorithm.

\begin{figure}[ht]
\centering
\begin{picture}(0,0)%
\includegraphics{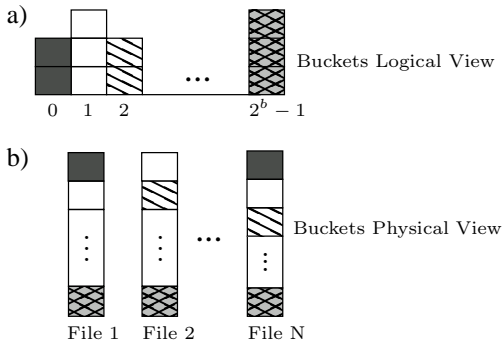}%
\end{picture}%
\setlength{\unitlength}{4144sp}%
\begingroup\makeatletter\ifx\SetFigFont\undefined%
\gdef\SetFigFont#1#2#3#4#5{%
  \reset@font\fontsize{#1}{#2pt}%
  \fontfamily{#3}\fontseries{#4}\fontshape{#5}%
  \selectfont}%
\fi\endgroup%
\begin{picture}(2924,2028)(1426,-7816)
\put(2332,-7200){\makebox(0,0)[lb]{\smash{{\SetFigFont{12}{14.4}{\familydefault}{\mddefault}{\updefault}.}}}}
\put(2332,-7268){\makebox(0,0)[lb]{\smash{{\SetFigFont{12}{14.4}{\familydefault}{\mddefault}{\updefault}.}}}}
\put(2332,-7335){\makebox(0,0)[lb]{\smash{{\SetFigFont{12}{14.4}{\familydefault}{\mddefault}{\updefault}.}}}}
\put(1892,-7200){\makebox(0,0)[lb]{\smash{{\SetFigFont{12}{14.4}{\familydefault}{\mddefault}{\updefault}.}}}}
\put(1892,-7268){\makebox(0,0)[lb]{\smash{{\SetFigFont{12}{14.4}{\familydefault}{\mddefault}{\updefault}.}}}}
\put(1892,-7335){\makebox(0,0)[lb]{\smash{{\SetFigFont{12}{14.4}{\familydefault}{\mddefault}{\updefault}.}}}}
\put(2962,-7304){\makebox(0,0)[lb]{\smash{{\SetFigFont{12}{14.4}{\familydefault}{\mddefault}{\updefault}.}}}}
\put(2962,-7349){\makebox(0,0)[lb]{\smash{{\SetFigFont{12}{14.4}{\familydefault}{\mddefault}{\updefault}.}}}}
\put(2962,-7394){\makebox(0,0)[lb]{\smash{{\SetFigFont{12}{14.4}{\familydefault}{\mddefault}{\updefault}.}}}}
\put(3151,-7171){\makebox(0,0)[lb]{\smash{{\SetFigFont{7}{8.4}{\familydefault}{\mddefault}{\updefault}Buckets Physical View}}}}
\put(1801,-7801){\makebox(0,0)[lb]{\smash{{\SetFigFont{7}{8.4}{\familydefault}{\mddefault}{\updefault}File 1}}}}
\put(2251,-7801){\makebox(0,0)[lb]{\smash{{\SetFigFont{7}{8.4}{\familydefault}{\mddefault}{\updefault}File 2}}}}
\put(2881,-7801){\makebox(0,0)[lb]{\smash{{\SetFigFont{7}{8.4}{\familydefault}{\mddefault}{\updefault}File N}}}}
\put(1683,-6466){\makebox(0,0)[lb]{\smash{{\SetFigFont{7}{8.4}{\familydefault}{\mddefault}{\updefault}0}}}}
\put(1895,-6466){\makebox(0,0)[lb]{\smash{{\SetFigFont{7}{8.4}{\familydefault}{\mddefault}{\updefault}1}}}}
\put(2109,-6466){\makebox(0,0)[lb]{\smash{{\SetFigFont{7}{8.4}{\familydefault}{\mddefault}{\updefault}2}}}}
\put(3173,-6171){\makebox(0,0)[lb]{\smash{{\SetFigFont{7}{8.4}{\familydefault}{\mddefault}{\updefault}Buckets Logical View}}}}
\put(2889,-6466){\makebox(0,0)[lb]{\smash{{\SetFigFont{7}{8.4}{\familydefault}{\mddefault}{\updefault}$2^b - 1$}}}}
\end{picture}%
\caption{Situation of the buckets at the end of the partitioning step: (a) Logical view (b) Physical view}
\label{fig:brz-partitioning}
\end{figure}

Figure~\ref{fig:brz-partitioning}(a) shows a \emph{logical} view of the
$2^b$ buckets generated in the partitioning step.
In reality, the 128-bit strings belonging to each bucket are distributed 
among many files,
as depicted in Figure~\ref{fig:brz-partitioning}(b).
In the example of Figure~\ref{fig:brz-partitioning}(b), 
the 128-bit strings in bucket 0 
appear in files 1 and $N$, the 128-bit strings in bucket 1 appear in files 1, 2
and $N$, and so on. 

This scattering of the 128-bit strings in the buckets could generate a performance
problem because of the potential number of seeks  
needed to read the 128-bit strings in each bucket from the $N$ files on disk 
during the second step. 
But, as we show later on in Section~\ref{sec:contr-disk-access}, the number of seeks 
can be kept small using buffering techniques.

The searching step is responsible for generating a MPHF for each bucket
and for computing the offset array.
Figure~\ref{fig:searchingstep} presents the searching step algorithm.
\begin{figure}[h]
\hrule 
\hrule 
\vspace{2.0mm}
\begin{tabbing}
aa\=type booleanx \==  (false, true); \kill
\> $\blacktriangleright$ Let $H$ be a minimum heap of size $N$, where the \\
\> ~~ order relation in $H$ is given by \\
\> ~~ $i = x[96,127] >> (32 - b)$ for $x\in F$ \\
\> $1.$ {\bf for} $j = 1$ {\bf to} $N$ {\bf do} \{ Heap construction \}\\
\> ~~ $1.1$ Read the first 128-bit string $x$ from File $j$ on disk\\
\> ~~ $1.2$ Insert $(i, j, x)$ in $H$ \\
\> $2.$ {\bf for} $i = 0$ {\bf to} $2^b - 1$ {\bf do} \\
\> ~~ $2.1$ Read bucket $B_i$ from disk driven by heap $H$ \\
\> ~~ $2.2$ Generate a MPHF for bucket $B_i$ \\
\> ~~ $2.3$ $\mathit{offset}[i+1] = \mathit{offset}[i] + |B_i|$ \\
\> ~~ $2.4$ Write the description of MPHF$_i$ and $\mathit{offset}[i]$ \\
\> ~~~~~~~ to the disk 
\end{tabbing}
\hrule 
\hrule 
\vspace{-1.0mm}
\caption{Searching step}
\label{fig:searchingstep}
\vspace{-3.0mm}
\end{figure}
Statement 1 of Figure~\ref{fig:searchingstep} constructs the heap $H$ of size $N$.
This is well known to be linear on $N$.
The order relation in $H$ is given by the bucket address $i$ 
(i.e., the $b$ most significant bits of $x \in F$).
Statement 2 has two important steps.
In statement 2.1, a bucket is read from disk,
as described below.
In statement 2.2, a MPHF is generated for each bucket $B_i$ using
the internal memory based algorithm presented 
in Section~\ref{sec:simple-mphf}.
In statement 2.3, the next entry of the offset array is computed. 
Finally, statement 2.4 writes the description of MPHF$_i$ and 
$\mathit{offset}[i]$ to disk. 
Note that to compute 
$\mathit{offset}[i+1]$ we just need the current bucket size and
$\mathit{offset}[i]$. So, we just need to maintain two entries
of vector $\mathit{offset}$ in memory all the time.

The algorithm to read bucket $B_i$ from disk is presented 
in Figure~\ref{fig:readingbucket}.
Bucket $B_i$ is distributed among many files and the heap $H$ is used to drive a
multiway merge operation.
Statement 1.1 extracts and removes triple 
$(i, j, x)$ from $H$, where $i$ is a minimum value in $H$.
Statement 1.2 inserts $x$ in bucket $B_i$.
Statement 1.3 performs a seek operation in File $j$ on disk for the first 
read operation and reads sequentially all 128-bit strings $x \in F$ 
that have the same index $i$ 
and inserts them all in bucket $B_i$.
Finally, statement 1.4 inserts in $H$ the triple $(i^\prime, j, x^\prime)$,  
where $x^\prime \in F$ is the first 128-bit string read from File $j$ (in statement 1.3) 
that does not have the same bucket address as the previous keys.


\begin{figure}[h]
\hrule 
\hrule
\vspace{2.0mm} 
\begin{tabbing}
aa\=type booleanx \==  (false, true); \kill
\> $1.$ {\bf while} bucket $B_i$ is not full {\bf do} \\
\> ~~ $1.1$ Remove $(i, j, x)$ from $H$\\
\> ~~ $1.2$ Insert $x$ into bucket $B_i$ \\
\> ~~ $1.3$ Read sequentially all 128-bit strings from File $j$\\
\> ~~~~~~~  that have the same $i$ and insert them into $B_i$ \\
\> ~~ $1.4$ Insert the triple $(i^\prime, j, x^\prime)$ in $H$, where $x^\prime$ is\\
\> ~~~~~~~ the first 128-bit string read from File $j$ that \\ 
\> ~~~~~~~ does not have the same bucket index $i$
\end{tabbing}
\hrule 
\hrule 
\vspace{-1.0mm}
\caption{Reading a bucket}
\label{fig:readingbucket}
\vspace{-1.0mm}
\end{figure}

It is not difficult to see from this presentation of the searching step 
that it runs in linear time.
To achieve this conclusion we use $O(N)$ computer words
to allow the merge operation to be performed in one pass through each file. 
In addition, it is also important to observe that:
\begin{enumerate}
\item $2^b < \frac{n}{\log n}$ (see Section~\ref{sec:hashfunctions}), 
\item $N \ll n$ (e.g., see Table~\ref{tab:diskaccess}
in Section~\ref{sec:contr-disk-access}) and
\item the internal algorithm runs in linear time, 
as shown in Section~\ref{sec:simple-mphf}.
\end{enumerate}

In conclusion,
our algorithm takes $O(n)$ time because both the partitioning and
the searching steps run in $O(n)$ time. 
The space required for constructing the resulting MPHF is 
$O(N)$ computer words because the memory usage in the partitioning
step does not depend on the number of keys in $S$ and, in the searching 
step, the internal algorithm is applied to problems of size up to 256. 
All together makes our algorithm the first one 
that demonstrates the capability of generating MPHFs for sets in the order
of billions of keys.

\subsection{The internal algorithm}
\label{sec:simple-mphf}

We now describe the internal algorithm, a simple and space-optimal 
way of constructing a minimal perfect hash function for a set
$S$ of $n$ elements.
We assume that we can create and access two truly random hash functions $f_0$ and $f_1$, mapping elements of $U$ to $\{0,\dots,\tau-1\}$, where $\tau \geq (1+\epsilon)n$ for some constant $\epsilon > 0$. We consider the bipartite graph $G$ with vertex set $V=\{0,\dots,2\tau-1\}$ and edge set $E=\{ \{f_0(x),\tau+f_1(x)\} \; | \; x\in S\}$. From the theory of random graphs~\cite{b01, hmwc93} we know that $G$ is acyclic with 
probability $\Omega(1)$, i.e, $Pr=e^{1/c}\sqrt{(c-2)/c}$, where $c = 2(1 + \epsilon)$. Thus, in an expected constant number of attempts we can find $f_0$ and $f_1$ such that $G$ is acyclic. In fact, this has been used in previous MPHF constructions~\cite{mwhc96}, but we will proceed differently to achieve a space usage of $O(n)$ bits rather than $O(n\log n)$ bits.

The data structure for the MPHF consists of two arrays of $2\tau$ bits, $T_1$ and $T_2$. The space usage for this is $4(1+\epsilon)n$ bits. We will use $T_1$ to associate a bit with every vertex of~$G$. For every connected component of $G$, which is a tree by choice of $f_0$ and $f_1$, we choose an arbitrary root node $v_0$ in $\{0,\dots,\tau-1\}$. Then, for a given non-isolated vertex $v$ of $G$ we can speak of its distance $d_v$ to the root of its component. We let $T_1[v]=1$ if $d_v\bmod 4 \in \{1,2\}$, and $T_1[v]=0$ otherwise. If $v$ is an isolated vertex, we let $T_1[v]=0$. It is easy to implement the initialization of $T_1$ in linear time by using a depth-first search algorithm. Now consider the following perfect hash function: 
{\small
\begin{equation*}
\phi(x)=
\begin{cases}
f_0(x) & \text{if $T_1[f_0(x)]\oplus T_1[\tau+f_1(x)] = 0$}\\
\tau+f_1(x) & \mbox{otherwise.}
\end{cases}
\end{equation*}
}(Again we denote exclusive-or by $\oplus$.) Now we will argue that the function $\phi$ is 1-1 on $S$. An element $x\in S$ corresponds to the edge $e=\{f_0(x),\tau+f_1(x)\}$ in $G$. We argue that 
$\phi(x)$ equals the vertex in $e$ that is furthest away from the root of the tree in which $e$ is part: This is clearly true for edges containing the root element, and the value of $T_1[f_0(x)]\oplus T_1[\tau+f_1(x)]$ changes at each step on a root-to-leaf path, as it should. Thus, $\phi$ is 1-1 on $S$.

$T_2$ is used to map down to the interval $\{0,\dots,n-1\}$ rather than $\{0,\dots,2\tau-1\}$. Specifically, we use $T_2$ to store the set $\phi(S)$ (indicating elements in the set by 1s). 
Note that $T_2$ can also be computed in linear time.
For a set $Y$ of integers, let rank$_Y(x)=| \{ y\in Y \;|\; y<x \} |$. Then the following is a minimal perfect hash function for $S$:
{\small
\begin{equation*}
h(x) = \mbox{rank}_{\phi(S)}(\phi(x))
\end{equation*}
}We need to specify how to compute the rank function, where the set is represented as a bit vector $T_2$. Note that computing rank($i$) corresponds to counting the number of 1s in $T_2$ in positions $0,\dots,i-1$. This is a well-studied primitive in succinct data structures, and it is known that it is possible to compute a rank in constant time, by using $o(\tau)$ additional space, see e.g.~\cite{dict-jour}. 

For completeness, we describe a practical variant that uses $\varepsilon \tau$ additional bits of space, where $\varepsilon$ can be chosen as any positive number. The evaluation time is $O(1/\varepsilon)$.
Conceptually, the scheme is very simple: Store explicitly the rank of every $\kappa$th index, where $\kappa =\lfloor\log(\tau)/\varepsilon\rfloor$. To compute rank($i$) we look up the rank of the largest precomputed index $j\leq i$, and count the number of 1s from position $j$ to $i-1$. To do this in time $O(1/\varepsilon)$ we use a lookup table that allows us to count the number of 1s in $\Omega(\log \tau)$ bits in constant time.
Such a lookup table takes $\tau^{\Omega(1)}$ space. Note that if, as in this paper, we have many MPHFs, they can all share a lookup table that may be larger than each individual MPHF, to reduce the constant in the rank computation.

\subsubsection{Improving the space}
\label{sec:simple-mphf-improved}

We now sketch a way of improving the space to just over 3 bits per key, adding a little complication to the scheme. We can notice that the contents of $T_1$ and $T_2$ are not independent. Specifically, there can be a non-zero bit in $T_1[i]$ only if $T_2[i]=1$. We can create a compressed representation $T'_1$ that uses only $n$ bits and enables us to compute any bit of $T_1$ in constant time. First of all, if $T_2[i]=0$ we can conclude that $T_1[i]=0$. We want to initialize $T'_1$ such that $T_1[i]=T'_1[\mbox{rank}_{\phi(S)}(i)]$ whenever $T_2[i]=1$, i.e., $i\in\phi(S)$. This is possible since $\mbox{rank}_{\phi(S)}(i)$ is 1-1 on elements in $\phi(S)$. In conclusion, we can replace $T_1$ by $T'_1$ and reduce the space usage to $2(1+\varepsilon)\tau+n$ bits. It is easy to note that $T'_1$ can be computed in linear time as well.

\subsubsection{The parameters choice for the internal algorithm}
The first parameter we are going to discuss is that $\epsilon$
responsible for allowing us to construct an acyclic random graph
with high probability. We have set
$\epsilon$ to $0.045$ in order to get a probability 
of approximately $0.33$ of generating a random graph with no cycles.
As a consequence the expected number of iterations to generate an acyclic 
graph is approximately 3, which comes from $1/Pr$.

The larger is the value of $\epsilon$, the 
sparser is the random graph used and, consequently,
the larger is the storage requirements of the resulting MPHFs and
the faster is the internal algorithm because of the greater probability 
of getting an acyclic random graph.
We have chosen a small value for $\epsilon$ because we are interested 
in more compact functions and the runtime of the internal algorithm is 
dominated by the time spent with I/Os.  

The parameter $\kappa$ is left to be set by the users so that they can trade
space for evaluation time and vice-versa.
In the experiments we
set $\kappa$ to 128 in order to spend less space to store the MPHF 
of each bucket.
This means that we store in a rank table the number of bits
set to 1 before every $128$th entry in the bit vector $T_2$.
As $\ell$ is upper bounded by $256$, then at most four and 
typically two 8-bit integers 
are required to store the rank values for each bucket.

To compute rank($i$) we look up the rank of the largest precomputed index $j\leq i$, and count the number of 1s from position $j$ to $i-1$. 
To do this we use a lookup table that allows us to count the number of 1s in $16$ bits in constant time. Therefore, to compute the number of bits set to 1 in 128 bits we need 
8 lookups.
Such a lookup table takes $2^{16}$ bytes of space that are shared for all the buckets. 
We could trade space for evaluation time by using a lookup table of $2^{8}$ bytes instead. 
However, $2^{16}$ bytes is small enough to fit in the cache and to have constant evaluation time.

\subsection{Hash functions used by the algorithms}
\label{sec:hashfunctions}
The aim of this section is threefold.
First, in Section~\ref{sec:definitions}, we define the hash function $h_0$ used to split the 
key set $S$ into $2^b$ buckets and, the hash functions $h_{i1}$ and
$h_{i2}$ used by the internal algorithm to generate the MPHF of each bucket,
where $0\leq i \leq 2^b-1$.
Second, in Section~\ref{sec:hash-functions}, we present the implementation 
details of those hash functions.
Third, in Section~\ref{sec:analysis-hash-functions}, we show the conditions that parameter 
$b$ must meet so that no bucket with more than $\ell$ keys is created by $h_0$.
We also show that $h_{i1}$ and $h_{i2}$ are truly random hash functions for 
the buckets.

\subsubsection{Definitions}
\label{sec:definitions}

We have made the design decision to make use of tabulation based hash functions, 
which seem to be a more practical alternative than hash functions based on 
integer multiplication of keys.\footnote{For example, as far as we know 
the best way of implementing multiplication of two 64-bit integers on 
contemporary machines is  by ``school method'' reduction to 4 multiplications 
of 32-bit integers. Similarly, a ``$\mod p$'' operation on the resulting 
128-bit integer, where $p$ is a 32-bit integer, seems to require 3 multiplications 
of 32-bit integers and 4 modulo operations on 64-bit integers.}
We will make extensive use of the linear hash functions analyzed by 
Alon, Dietzfelbinger, Miltersen and Petrank~\cite{admp99}. 
To do so we consider a key as a 0-1 vector of length $L$. 
The variable-length strings that we consider are conceptually made fixed 
length by padding with zeros at the end, which results in a unique vector 
since ascii character 0 does not appear in any string. 

Mathematically, $h_0$ is a randomly chosen linear map over GF(2) from $\{0,1\}^L$ to $\{0,1\}^b$. To get an efficient implementation, we use a tabulation idea from~\cite{an96} where we can get evaluation time $O(L/\log \sigma)$ by using space $L \sigma$ -- see Section~\ref{sec:hash-functions} for implementation details. Choosing $\sigma = n^{\Omega(1)}$ we obtain evaluation time $O(L/\log n)$. (In theory we could get evaluation time $O(L/w)$, where $w\geq \log n$ is the word length of the computer, by first hashing down to $O(\log n)$ bits using universal hashing; however, this does not seem to give an improvement in practice.) We choose $b$ as small as possible such that the maximum bucket size is bounded by $\ell$ with reasonable probability (some constant close to 1). 
By a result of~\cite{admp99} we know that 
\begin{equation} \label{eq:parameterb}
b\leq \log n-\log(\ell/\log\ell) + O(1).
\end{equation} 
For the implementation, we will experimentally determine the smallest possible choices of $b$.

To define $h_{i1}$ and $h_{i2}$ we proceed as follows. Again use the linear hash function 
of~\cite{admp99} to implement hash functions $y_1,\dots,y_k$ from $\{0,1\}^L$ to
$\{0,1\}^{r-1}0$, where $r\gg \log\ell$ and $k$ are parameters to be determined later. 
Note that the range is the set of $r$-bit strings ending with a $0$. 
The purpose of the last 0 is to ensure that we can have no collision  
between $y_j(x_1)$ and $y_j(x_2)\oplus 1$, $1 \leq j \leq k$,
for any pair of elements $x_1$ and $x_2$.
Let $p$ be a prime number much larger than the size of the desired range of $h_{i1}$ and $h_{i2}$, which in our case is $|B_i|$, 
and let $t_1,\dots,t_{2k}$ be tables of $2^r$ random values in $\{0,\dots,p-1\}$. We then define:
{\small
\begin{align}\label{eq:hi1hi2}
\rho(x,s,\Delta) & =\left(\sum_{j=1}^{k} t_j[y_j(x) \oplus \Delta] + s\sum_{j=k+1}^{2k}t_j[y_j(x) \oplus\Delta]\right) \bmod p \nonumber\\
h_{i1}(x) & =\rho(x,s_i,0) \bmod |B_i| \nonumber \\
h_{i2}(x) & =\rho(x,s_i,1) \bmod |B_i|
\end{align}
}where the symbol $\oplus$ denotes exclusive-or and the variable $s_i$ is specific to bucket $i$. To find $s_i$ we choose random values from $\{1,\dots,p-1\}$ until the functions $h_{i1}$ and $h_{i2}$ work with the internal algorithm of Section~\ref{sec:simple-mphf}. It is known that a constant fraction of the set of all functions work; in Section~\ref{sec:analysis-hash-functions} we will argue that this will also be the case when the hash functions are chosen as above.

\subsubsection{Implementation details}
\label{sec:hash-functions}
In order to implement the functions $h_0$, $y_1, y_2, y_3, \dots, y_k$ to be computed 
at once we use a function $h'$ from a family of
linear hash functions over GF(2) proposed by Alon, Dietzfelbinger, Miltersen and 
Petrank~\cite{admp99}.
The function has the following form: $h'(x)=Ax$, where $x \in S$ and $A$ is 
a $\gamma \times L$ matrix in which the elements are randomly chosen from $\{0,1\}$.
The output is a bit string of an a priori defined size $\gamma$. 
In our implementation $\gamma = 128$ bits. 
It is important to realize that this is a matrix multiplication over GF (2). 
The implementation can be done using a bitwise-and operator (\&) and 
a function $f:\{0,1\}^\gamma\to \{0,1\}$ to compute parity instead of 
multiplying numbers. 
The parity function $f(a)$ produces 1 as a result
if $a \in \{0,1\}^\gamma$ has an odd number of bits set to 1,
otherwise the result is 0.
For example, let us consider $L=3\:\mathrm{bits}$, 
$\gamma=3\:\mathrm{bits}$, $x=110$ and 
{\small
\[ A = \left[ \begin{array}{ccc}
1 & 0 & 1 \\
0 & 0 & 1 \\
1 & 1 & 0 \end{array} \right]\cdot\] 
}The number of rows gives the required number of bits in the output, i.e., $\gamma=3$.
The number of columns corresponds to the value of $L$. Then,
{\small
\[ h'(x) = \left[ \begin{array}{ccc}
1 & 0 & 1 \\
0 & 0 & 1 \\
1 & 1 & 0 \end{array} \right] 
\left[ \begin{array}{c}
1 \\
1 \\
0 \end{array} \right] = 
\left[ \begin{array}{c}
b_1 \\
b_2 \\
b_3 \end{array} \right]
\]
}where $b_1 = f(101 \:\: \& \: 110) = 1$, 
$b_2 = f(001 \:\: \& \: 110) = 0$ and
$b_3 = f(110 \:\: \& \: 110) = 0$.

To get a fast evaluation time, some tabulation is  required.
Note that if $x$ is short, e.g. 8 bits, we can simply tabulate all the  function values
and compute $h'(x)$ by looking up the value $h'[x]$ in an array $h'$.
To make the same thing work for longer keys, split the matrix $A$ into  parts of
$8$ columns each: $A = A_1 | A_2 | \dots | A_{\lceil L/8 \rceil}$, and create a lookup table  $h'_{i}$ for each
submatrix. Similarly split $x$ into parts of $8$ bits, $x = x_1 x_2 \dots x_{\lceil L/8 \rceil}$.
Now $h'(x)$ is the exclusive-or of $h'_{i}[x_i]$, for $i=1,\dots, \lceil L/8 \rceil$.
Therefore, we have set $\sigma$ to 256 so that keys of size $L$ can be processed 
in chunks of $\log \sigma =8$ bits. In our URL collection the largest key 
has 65 bytes, i.e., $L = 520$ bits.

The 32 most significant bits of $h'(x)$, where $x \in S$, are used to compute
the bucket address of $x$, i.e., $h_0(x) = h'(x)[96, 127] >> (32 -b)$.
We use the symbol $>>$ to denote the right shift of bits.
The other 96 bits correspond to $y_1(x), y_2(x), \dots y_6(x)$, taking $k = 6$.
This would give $r = 16$, however, to save space for storing 
the tables used for computing $h_{i1}$ and $h_{i2}$, 
we hard coded the linear hash function
to make the most and the least significant bit of each chunk of 16 bits 
equal to zero.  Therefore, $r=15$. 
This setup enable us to solving problems 
of up to 500 billion keys, which is plenty of 
for all the applications we know of. 
If our algorithm fails in any phase, we just restart it. 
As the parameters are chosen to have success with high probability, the 
number of reinitializations is $O(1)$. 

Finally, the last parameter related to the hash functions we need to talk about
is the prime number $p$. As $p$ must be much larger than 
the range of $h_{i1}$ and $h_{i2}$, then we set it to
the largest 32-bit integer that is a prime, i.e, $p = 4294967291$.

\subsubsection{Analytical results}
\label{sec:analysis-hash-functions}

In this section we sketch the analysis of the hash functions of our scheme. Note that the hash functions $h_0$, $y_1$, $y_2$, $\dots$, $y_k$ have a range of $b+kr$ bits in total. Thus, by universality of linear hash functions~\cite{admp99}, the probability that there exist two keys that have the same values under all functions is at most $\binom{n}{2}/2^{b+kr}$. We will choose $r$ such that this probability becomes negligible.
For simplicity, we assume that the zero vector $0^L$ is not in the set $S$ -- it is not hard to see that this assumption is insignificant.

A direct consequence of Theorem~5 in~\cite{admp99} is that, assuming $b\leq\log n-\log\log n$, the expected size of the largest bucket is $O(n\log b / 2^b)$, i.e., a factor $O(\log b)$ from the average 
bucket size. This justifies the choice of $b$ in Eq.~(\ref{eq:parameterb}), imposing the requirement that
$\ell \geq \log n\log\log n$. 


For any choice of $s$, we will now analyze the probability (over the choice of  
$y_1,\dots,y_k$) that $x\mapsto \rho(x,s,0)$ and $x\mapsto \rho(x,s,1)$ map the 
elements of $B_i$ uniformly and independently to $\{0,\dots,p-1\}$. A sufficient
 criterion for this is that the sums $\sum_{j=1}^{k} t_j[y_j(x) \oplus \Delta]$ 
and $\sum_{j=k+1}^{2k}t_j[y_j(x) \oplus\Delta]$, $\Delta\in\{0,1\}$, have values
 that are uniform in $\{0,\dots,p-1\}$ and independent. This is the case if for 
every $x\in B_i$ there exists an index $j_x$ such that neither $y_{j_x}$ or 
$y_{j_x}\oplus 1$ belongs to $y_{j_x}(B_i - \{x\})$. Since $y_1,\dots,y_k$ are 
universal hash functions, the probability that this is not the case for a given element 
$x\in B_i$ is bounded by $(|B_i|/2^r)^k \leq (\ell/2^r)^k$. If we choose, for example 
$r=\lceil\log(\sqrt[3]{n}\ell)\rceil$
and $k=4$ we have that this probability is $o(1/n)$. Hence, the probability that
this happens for {\em any} key in $S$ is $o(1)$.

Finally, we need to argue that for each bucket $i$ it is easy to find a value of $s$ such that
the pair $h_{i1}$, $h_{i2}$ is good for the MPHF of the bucket. We know that with constant
probability this is the case if the functions were truly random. Now, as argued above, with
probability $1-o(1)$ the functions $x\mapsto \rho(x,s,0)$ and $x\mapsto \rho(x,s,1)$ are
random and independent on each bucket, for every value of $s$. Then, for a given bucket
and a given value of $s$ there is a probability $\Omega(1)$ that the pair of hash functions
work for that bucket. Now, for any $\Delta\in\{0,1\}$ and $s\neq s'$, the functions $x\mapsto \rho(x,s,\Delta)$ and $x\mapsto \rho(x,s',\Delta)$ are independent. Thus, by Chebychev's inequality
the probability that less than a constant fraction of the values of $s$ work for a given bucket is
$O(1/p)$. So with probability $1-o(1)$ there is a constant fraction of ``good'' choices of $s$ in
every bucket, which means that trying an expected constant number of random values for $s$
is sufficient in each bucket.


\section{Experimental results}
\label{sec:experimental-results}
In this section we present the experimental results.
We start presenting the experimental setup. 
We then present the performance of our algorithms 
considering construction time, storage space and evaluation
time as metrics for the resulting MPHFs.
Finally, we discuss how the amount of internal memory available 
affects the runtime of our two-step external memory based 
algorithm. 
\begin{table*}[htb]
\begin{center}
{\scriptsize
\begin{tabular}{|l|c|c|c|c|c|}
\hline
$n$ (millions)   & 1                 & 2                & 4                  & 8                   & 16             \\
\hline 
Average time (s) & $3.34 \pm 0.02$   & $6.97 \pm 0.02$  & $14.64 \pm 0.04$   & $31.75 \pm 0.49$    & $68.98 \pm 0.82$  \\
SD               & $0.03$            & $0.03$           & $0.05$             & $0.73$              & $1.22$         \\
\hline
\hline
$n$ (millions)   & 32                & 64               & 128                & 512                 & 1024            \\
\hline 
Average time (s) & $142.71 \pm 1.44$ & $288.95 \pm 2.65$& $604.70  \pm 6.22$ & $2383.08 \pm 22.11$ & $4982.97 \pm 55.14$  \\
SD               & $2.01$            & $3.70$           & $8.69$             & $28.77$             & $51.12$            \\
\hline

\end{tabular}
\vspace{-3mm}
}
\end{center}
\caption{External algorithm: average time in seconds for constructing a MPHF,
the standard deviation (SD), and the confidence intervals considering 
a confidence level of $95\%$.
}
\label{tab:mediasbrz}
\end{table*}

\subsection{The data and the experimental setup}
\label{sec:data-exper-set}

The algorithms were implemented in the C language and
are available at 
\texttt{http://\-www.dcc.ufmg.br/\texttt{\~ }fbotelho}
under the GNU Lesser General Public License (LGPL).
All experiments were carried out on
a computer running the Linux operating system, version 2.6,
with a 1 gigahertz  AMD Athlon 64 Processor 3200+
and 1 gigabyte of main memory. 

Our data consists of a collection of approximately 1 billion
URLs collected from the Web, each URL 64 characters long on average.
The collection is stored on disk in 60.5 gigabytes of space.

\subsection{Performance of the algorithms}
\label{sec:performance}

We are firstly interested in verifying the claim that 
our two-step external memory based algorithm runs in linear time.  
Therefore, we run the algorithm for several numbers $n$ of keys in $S$.
The values chosen for $n$ were 1, 2, 4, 8, 16, 32, 64, 128, 512 and $1024$
million.
We limited the main memory in 512 megabytes for the experiments
in order to show that the algorithm does not need much internal 
memory to generate MPHFs.
The size $\mu$ of the a priori reserved internal memory area 
was set to 200 megabytes. 
In Section~\ref{sec:contr-disk-access} we show how $\mu$
affects the runtime of the algorithm. 
The parameter $b$ (see Eq.~(\ref{eq:parameterb})) was set 
to the minimum value that gives us a maximum bucket size lower
than $\ell = 256$.
For each value chosen for $n$, the respective values for $b$ 
are $13, 14, 15, 16,17,18,19,20,22$ and $23$ bits.

In order to estimate the number of trials for each value of $n$ we use
a statistical method for determining a suitable sample size (see, e.g.,
\cite[Chapter 13]{j91}).  
We got that just one trial for each $n$ would be enough with a confidence level of $95\%$.
However, we made 10 trials.  This number of trials seems rather small, but, as
shown below, the behavior of our external algorithm is very stable and its runtime is
almost deterministic (i.e., the standard deviation is very small) because
it is a random variable that follows a
(highly concentrated) normal distribution.

Table~\ref{tab:mediasbrz} presents the runtime average for each $n$,
the respective standard deviations, and 
the respective confidence intervals given by 
the average time $\pm$ the distance from average time
considering a confidence level of $95\%$.
Observing the runtime averages we noticed that 
the algorithm runs in expected linear time, 
as we have claimed.  Better still,
it outputs the resulting MPHF faster than all practical algorithms we
know of, because of the following reasons. 
First, the memory accesses during the generation of a MPHF for
a given bucket cause cache hits, once the problem was broken down
into problems of size up to 256. 
Second, at searching step we are dealing with 16-byte (128-bit) strings 
instead of 64-byte URLs.



\begin{figure}[htb]
\begin{center}
\scalebox{0.5}{\includegraphics{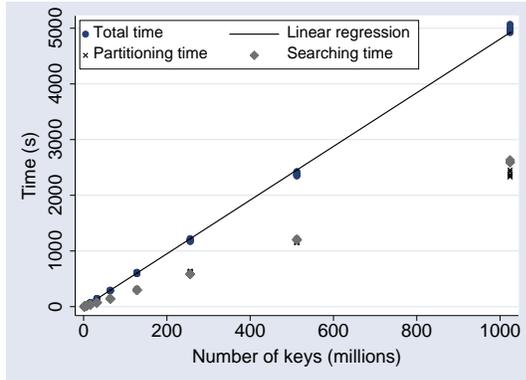}}
\caption{Partitioning time and searching time versus number of keys in $S$ for our external algorithm. 
The solid line corresponds to a linear regression model for the total time.}
\label{fig:brz_temporegressao}
\end{center}
\end{figure}

Figure~\ref{fig:brz_temporegressao}
presents the runtime for each trial. In addition, 
the solid line corresponds to a linear regression model 
obtained from the experimental measurements.
As we were expecting the runtime for a given $n$ has almost no 
variation.
The percentages of the total time spent in the partitioning step and in the searching are
approximately $49\%$ and $51\%$, respectively.

An intriguing observation is that the runtime of the algorithm is almost
deterministic, in spite of the fact that it uses as building block an
algorithm with a considerable fluctuation in its runtime.  
A given bucket~$i$,
$0 \leq i < 2^b$, is a small set of keys (at most 256 keys) and,
the runtime of the
building block algorithm is a random variable~$X_i$ with high fluctuation
(it follows a geometric distribution with mean~$1/Pr \approx 3$).
However, the runtime~$Y$ of the searching step of our external algorithm is given
by~$Y=\sum_{0\leq i< 2^b}X_i$.  Under the hypothesis that
the~$X_i$ are independent and bounded, the {\it law of large numbers} (see,
e.g., \cite{j91}) implies that the random variable $Y/2^b$
converges to a constant as~$n\to\infty$.  
This explains why the runtime is almost deterministic.

The next important metric on MPHFs is the space required to 
store the functions.
In order to apply the internal algorithm to
larger sets we randomly choose $f_0$ and $f_1$ 
from the family of universal hash functions
proposed by Thorup~\cite{t00}.
The internal algorithm was analyzed under the full 
randomness assumption so that universal hashing is
not enough to guarantee that it works for every key set.
But it has been the case for every key set we have applied it to.
Then, we refer to this version as {\it heuristic internal algorithm}.

Table~\ref{tab:bitsperkey} shows
how many bits per key the heuristic internal algorithm requires to 
store the resulting MPHFs.
In our setup the heuristic internal algorithm requires around $2.1$ and $3.3$ bits per key 
to respectively store the resulting PHFs and MPHFs.
In a PC with 1 gigabyte of main memory the largest set we are able to generate a 
MPHF for is a set with 30 millions of keys,
because of the sparse graph required to generate the functions
is memory demanding. 


\begin{table}[ht]
\begin{center}
{\scriptsize
\begin{tabular}{|c|c|c|}
\hline
\raisebox{-0.7em}{$n$}    & \multicolumn{2}{c|}{\raisebox{-1mm}{Bits/key}} \\
\cline{2-3}
                          & \raisebox{-0.4mm}{PHF} &  \raisebox{-0.4mm}{MPHF} \\
\hline
$10^4$                    & $2.13$ &  $3.37$                                  \\
$10^5$                    & $2.09$ &  $3.32$                                  \\
$10^6$                    & $2.09$ &  $3.32$                                  \\
$10^7$                    & $2.09$ &  $3.32$                                  \\
\hline
\end{tabular}
}
\end{center}
\caption{Heuristic internal algorithm: space usage to respectively store the resulting PHFs and MPHFs.}
\label{tab:bitsperkey}
\end{table}

The external algorithm is designed to be used when the 
key set does not fit in main memory. 
Table~\ref{tab:bitsperkeyext} shows that 
it can be used for constructing 
PHFs and MPHFs that require approximately 2.7 and 3.8 bits 
per key to be stored, respectively. 
The lookup tables used by the hash functions of the external algorithm 
require a fixed storage cost of 1,847,424 bytes. 
This makes the external algorithm 
unsuitable for sets with less than 16 million of keys.

\begin{table}[ht]
\begin{center}
{\scriptsize
\begin{tabular}{|c|c|c|c|c|}
\hline
\raisebox{-0.7em}{$n$}    & \raisebox{-0.7em}{$b$} &  \raisebox{-0.3em}{Cost in bytes to } &\multicolumn{2}{c|}{\raisebox{-1mm}{Bits/key}}\\
\cline{4-5}
                          &                        &  \raisebox{0.3em}{store lookup tables} &   \raisebox{-0.4mm}{PHF}      & \raisebox{-0.4mm}{MPHF}\\
\hline
$10^4$                    &       6                &  1,847,424  & $2.93$  & $3.71$     \\
$10^5$                    &       9                &  1,847,424  & $2.73$  & $3.57$     \\
$10^6$                    &      13                &  1,847,424  & $2.65$  & $3.82$     \\
$10^7$                    &      16                &  1,847,424  & $2.51$  & $3.70$     \\
$10^8$                    &      20                &  1,847,424  & $2.80$  & $4.02$     \\
$10^9$                    &      23                &  1,847,424  & $2.65$  & $3.83$     \\
\hline
\end{tabular}
}
\end{center}
\caption{External algorithm: space usage to respectively store the resulting PHFs and MPHFs.}
\label{tab:bitsperkeyext}
\end{table}

To overcome the problem mentioned above we have implemented a version of the external 
algorithm that uses the pseudo random hash function proposed by
Jenkins~\cite{j97}.
This function was used instead of the linear hash function
described in Section~\ref{sec:hash-functions},
and instead of the two 
truly random hash function of each bucket, i.e.,
$h_{i1}$ and $h_{i2}$, where $0 \leq i < 2^b$.
This version is, from now on, referred to as 
{\it heuristic external algorithm}.
The Jenkins function just loops around the key 
doing bitwise operations over chunks of 12 bytes.
Then, it returns the last chunk. Thus, in the mapping 
step, the key set S is mapped to F, which now 
contains 12-byte long strings instead of 
16-byte long strings. 

The Jenkins function needs just one random seed of 
32 bits to be stored instead of quite long lookup tables.
Therefore, there is no fixed cost to store the resulting 
MPHFs, but two random seeds of 32 bits
are required to describe the functions $h_{i1}$ and $h_{i2}$
of each bucket.
As a consequence, the MPHFs generation 
and the MPHFs efficiency at retrieval time are 
faster (see Table~\ref{tab:constructiontime} and \ref{tab:evaltime}).
The reasons are twofold. 
First, we are dealing with 12-byte strings instead of
16-byte strings. 
Second, there are no large lookup tables to 
cause {\it cache misses}. For example, the construction 
time for a set of 1024 million keys has dropped down to $1.04$ hours
in the same setup.
The disadvantage of using the Jenkins function is that 
there is no formal proof that
it works for every key set. 
That is why the hash functions we have designed in this 
paper are required, even being slower. 
In the implementation available, the hash functions to be used
can be chosen by the user.
 
Table~\ref{tab:constructiontime} presents a comparison 
of our methods with the ones proposed by
Pagh~\cite{p99} (Hash-displace),
by Botelho, Kohayakawa and Ziviani~\cite{bkz05} (BKZ),
by Czech, Havas and Majewski~\cite{chm92} (CHM), and
by Fox, Chen and Heath~\cite{fch92} (FCH),
considering construction time and storage space
as metrics. 
The form of the MPHFs generated by those methods 
is presented in Section~\ref{sec:relatedprevious-work}.
Notice that they are the most important practical results on MPHFs
known in the literature.
Observing the results, our heuristic internal algorithm is the best choice
for sets that can be handled in main memory and our external 
algorithm is the first one that can be applied to sets 
that do not fit in main memory.

\begin{table}[ht]
\begin{center}
{\scriptsize
\begin{tabular}{|c|c|c|c|}
\hline
\multicolumn{4}{|c|}{Time in seconds to construct a MPHF for $2\times10^6$ keys} \\
\hline
\raisebox{-0.7em}{Algorithms}&  \raisebox{-0.4em}{Function} &  \raisebox{-0.4em}{Construction}      & \raisebox{-0.7em}{bits/key} \\
                             &  \raisebox{0.4em}{type}      &  \raisebox{0.4em}{time (seconds)}     &          \\
\hline
Heuristic Internal  & PHF   & $12.99 \pm 1.01$             &  $2.09$   \\
\cline{2-4}
Algorithm           & MPHF  & $13.94 \pm 1.06$             &  $3.35$   \\
\hline
External            & PHF   & $6.92  \pm 0.04$             &  $2.64$  \\
\cline{2-4}
Algorithm           & MPHF  & $6.98  \pm 0.01$             &  $3.85$  \\
\hline
\raisebox{-0.3em}{Heuristic External}  & \raisebox{-0.7em}{MPHF}& \raisebox{-0.7em}{$4.75  \pm 0.02$} & \raisebox{-0.7em}{$3.7$}  \\
 \raisebox{0.3em}{Algorithm}           &       &                            &   \\
\hline
Hash-displace       & MPHF  & $46.18 \pm 1.06$             &  $64.00$  \\
BKZ                 & MPHF  & $8.57  \pm 0.31$             &  $32.00$  \\
CHM                 & MPHF  & $13.80 \pm 0.70$             &  $66.88$  \\
FCH                 & MPHF  & $758.66 \pm 126.72$          &  $5.84$   \\
\hline
\end{tabular}
}
\end{center}
\caption{Construction time and storage space without considering the fixed 
cost to store lookup tables.}
\label{tab:constructiontime}
\end{table}

Finally, we show how efficient is the
resulting MPHFs at retrieval time for the methods aforementioned,
which is as important as construction time and storage space.
Table~\ref{tab:evaltime} presents the time, in seconds, to
evaluate $2\times10^6$ keys. 
We group the BKZ and CHM methods together because the resulting 
MPHFs have the same form. 
From the results we can conclude that our heuristic internal algorithm
generates MPHFs that are as fast to be computed as
the ones generated by the most practical methods on MPHFs.
The MPHFs generated by the external algorithm are slower. 
Nevertheless, the difference is not so expressive (each key can be evaluated 
in few microseconds) and the external algorithm is the first efficient option 
for sets that do not fit in main memory. 

\begin{table}[ht]
\begin{center}
{\scriptsize
\begin{tabular}{|c|c|c|c|c|c|c|}
\hline
\multicolumn{7}{|c|}{Time in seconds to evaluate $2\times10^6$ keys} \\
\hline
\raisebox{-0.7em}{key length (bytes)}  &  \raisebox{-0.4em}{Function}  &  \raisebox{-0.7em}{8}     & \raisebox{-0.7em}{16}     & \raisebox{-0.7em}{32}     & \raisebox{-0.7em}{64}     & \raisebox{-0.7em}{128} \\
                                       &  \raisebox{0.4em}{type}       &        &        &        &        &     \\
\hline
Heuristic Internal  & PHF     & 0.41 & 0.55 & 0.79 & 1.29 & 2.39 \\
\cline{2-7}
Algorithm           & MPHF    & 0.85 & 0.99 & 1.23 & 1.73 & 2.74 \\
\hline
External            & PHF     & 2.05 & 2.31 & 2.84 & 3.99 & 7.22 \\
\cline{2-7}
Algorithm           & MPHF    & 2.55 & 2.83 & 3.38 & 4.63 & 8.18 \\
\hline
\raisebox{-0.3em}{Heuristic External}& \raisebox{-0.7em}{MPHF}& \raisebox{-0.7em}{1.19} & \raisebox{-0.7em}{1.35} & \raisebox{-0.7em}{1.59} & \raisebox{-0.7em}{2.11} & \raisebox{-0.7em}{3.34} \\
\raisebox{0.3em}{Algorithm}          &                        &      &      &      &      &  \\
\hline
Hash-displace       & MPHF   & 0.56 & 0.69 & 0.93 & 1.44 & 2.54 \\
BKZ/CHM             & MPHF   & 0.61 & 0.74 & 0.98 & 1.48 & 2.58 \\
FCH                 & MPHF   & 0.58 & 0.72 & 0.96 & 1.46 & 2.56 \\
\hline
\end{tabular}
}
\end{center}
\caption{Evaluation time.}
\label{tab:evaltime}
\end{table}

It is important to emphasize that 
the BKZ, CHM and FCH methods were analyzed under the full randomness assumption
as well as our heuristic internal algorithm. Therefore, 
our external algorithm is the first one that has
experimentally proven practicality for large key sets
and has both space usage 
for representing the resulting functions and the construction time carefully proven.
Additionally, it is the fastest algorithm for constructing the functions and 
the resulting functions are much simpler than the ones 
generated by previous theoretical well-founded schemes 
so that they can be used in practice. Also, it
considerably improves the first step given by Pagh with 
his hash and displace method~\cite{p99}.

\subsection{Controlling disk accesses}
\label{sec:contr-disk-access}

In order to bring down the number of seek operations on disk
we benefit from the fact that our external algorithm leaves almost all main
memory available to be used as disk I/O buffer. 
In this section we evaluate how much the parameter $\mu$ 
affects the runtime of our external algorithm.
For that we fixed $n$ in approximately 1 billion of URLs,
set the main memory of the machine used for the experiments 
to 1 gigabyte and used $\mu$ equal to 100, 200, 300, 400 and 500
megabytes. 

In order to amortize the number of seeks performed we use a buffering technique~\cite{k73}.
We create a buffer $j$ of size \textbaht$\: = \mu/N$ for each file $j$, 
where $1\leq j \leq N$.
Every time a read operation is requested to file $j$ and the data is not found 
in the $j$th~buffer, \textbaht~bytes are read from file $j$ to buffer $j$. 
Hence, the number of seeks performed in the worst case is given by
$\beta/$\textbaht~(remember that $\beta$ is the size in bytes of the fixed-length 
key set $F$).
For that we have made the pessimistic assumption that one seek happens every time 
buffer $j$ is filled in. 
Thus, the number of seeks performed in the worst case is $16n/$\textbaht, since
after the partitioning step we are dealing with 128-bit (16-byte) strings instead of 
64-byte URLs, on average. Therefore, the number of seeks is linear on 
$n$ and amortized by \textbaht.
It is important to emphasize that the operating system uses techniques
to diminish the number of seeks and the average seek time. 
This makes the amortization factor to be greater than \textbaht~in practice.  

Table~\ref{tab:diskaccess} presents the number of files $N$,
the buffer size used for all files, the number of seeks in the worst case considering
the pessimistic assumption aforementioned, and 
the time to generate a (minimal)PHF for approximately 1 billion of keys as a function of the amount of internal 
memory available. Observing Table~\ref{tab:diskaccess} we noticed that the time spent in the construction
decreases as the value of $\mu$ increases. However, for $\mu > 400$, the variation 
on the time is not as significant as for $\mu \leq 400$. 
This can be explained by the fact that the kernel 2.6 I/O scheduler of Linux
has smart policies  
for avoiding seeks and diminishing the average seek time 
(see \texttt{http://www.linuxjournal.com/article/6931}).
\begin{table}[ht]
\begin{center}
{\scriptsize
\begin{tabular}{|l|c|c|c|c|c|}
\hline
$\mu$ (MB)                                                        & $100$                        & $200$                       & $300$                       & $400$                       & $500$                       \\
\hline
$N$ (files)                                                       & $245$                        & $99$                        & $63$                        & $46$                        & $36$                        \\
\textbaht~(in KB)                                     & $418$                        & $2,069$                     & $4,877$                     & $8,905$                     & $14,223$                    \\
$\beta$/\textbaht                                                 & $151,768$                    & $30,662$                    & $13,008$                    & $7,124$                    & $4,461$                      \\
Time (hours)                                                      & $1.58$                       & $1.37$                      & $1.33$                        & $1.32$                      & $1.32$                    \\
\hline
\end{tabular}
}
\end{center}
\caption{Influence of the internal memory area size ($\mu$) in our external algorithm runtime.}
\label{tab:diskaccess}
\end{table}

\section{Concluding remarks}
\label{sec:concuding-remarks}

This paper has presented two novel algorithms for 
constructing PHFs and MPHFs and three implementations 
of the algorithms. The implementations in the C language 
are available at \texttt{http://anonymous}
under the GNU Lesser General Public License (LGPL).

The first algorithm, referred to as internal algorithm,
assumes that two truly random hash functions $f_0$ and $f_1$ are available
for free so that a PHF or a MPHF can be constructed from the acyclic random 
graph induced by $f_0$ and $f_1$. 
The full randomness assumption is not realistic because 
each truly random hash functions would require at least $n\log n$
bits to be stored, which is memory demanding.

In order to compare the internal algorithm with the 
most important practical results on 
MPHFs that
consider the same assumption (see Section~\ref{sec:relatedprevious-work}) 
we have 
chosen the required hash functions from the 
family of universal hash functions proposed by Thorup~\cite{t00}.
As universal hash functions are not enough to guarantee that
the algorithm would work for every key set, then we have referred to
this version of the algorithm as heuristic internal algorithm.

The heuristic internal algorithm outperforms all 
previous methods considering the storage space required for
the resulting functions.  
The resulting PHFs and MPHFs require
approximately 2.1 and 3.3 bits per key to be stored,
respectively.
Better still, the resulting functions are almost as fast 
to be computed and generated as the ones coming from
previous methods known in the literature. 
Tables~\ref{tab:bitsperkey},
\ref{tab:constructiontime} and \ref{tab:evaltime}
summarize the experimental results.

The second algorithm, referred to as external algorithm,
contains, as a component, a provably good implementation of 
the internal memory algorithm.
This means that the two hash functions $h_{i1}$ and
$h_{i2}$ (see Eq.~(\ref{eq:hi1hi2})) used instead of 
$f_0$ and $f_1$ behave as truly random hash functions
(see Section~\ref{sec:analysis-hash-functions}).
The resulting PHFs and MPHFs require approximately 2.7 and 3.8 bits 
per key to be stored and are generated faster than the ones generated
by all previous methods (including our heuristic internal algorithm).
The external algorithm is the first one that 
has experimentally proven practicality for sets in the order 
of billions of keys and 
has time and space usage carefully analyzed without unrealistic 
assumptions.
As a consequence, the external algorithm will work for every
key set. 

The resulting functions of the external algorithm are approximately 
four times slower than
the ones generated by our heuristic internal algorithm and by
all previous practical methods (see Table~\ref{tab:evaltime}).
The reason is that to compute the involved hash functions
we need to access lookup tables that do not fit in the cache.
To overcome this problem, at the expense of losing the 
guarantee that it works for every key set, we have 
proposed a heuristic version of the external algorithm that
uses a very efficient pseudo random hash function proposed by 
Jenkins~\cite{j97}. The resulting functions require 
the same storage space, are now less than  two times slower to be computed 
and are still faster to be generated.

Besides the data management applications for minimal 
perfect hash functions described in Section~\ref{sec:intro}, the 
external algorithm will be very useful for the information 
retrieval community as well. Search engines are nowadays indexing 
tens of billions of pages and the work with huge collections is 
becoming a daily task. For instance, the simple assignment of 
number identifiers to web pages of a collection can be a challenging 
task. While traditional databases simply cannot handle more traffic 
once the working set of URLs does not fit in main memory 
anymore~\cite{s05}, the external algorithm we propose here to 
construct MPHFs can easily scale to billions of entries. 
Also, algorithms like PageRank~\cite{Brin1998}, which uses the web 
link structure to derive a measure of popularity for Web pages, 
operates on the web graph. At construction time of the graph,
the URLs must be mapped to integers that will be used to label 
the vertices.
For the same reason, the WebGraph research group~\cite{bv04} 
would also benefit from a MPHF for sets in the order of billions 
of URLs to scale and to improve the storage requirements of their 
algorithms on Graph compression.



\bibliographystyle{abbrv}

\balancecolumns

\end{document}